\documentclass[%
 aip,
 amsmath,amssymb,
 reprint,%
]{revtex4-1}

\usepackage{graphicx}
\usepackage{dcolumn}
\usepackage{bm}
\usepackage[utf8]{inputenc}
\usepackage[T1]{fontenc}
\usepackage{mathptmx}
\usepackage{amsmath}
\usepackage{array}
\usepackage{booktabs}
\usepackage{multirow}
\usepackage{float}
\usepackage{xcolor}
\usepackage{changes}
\usepackage[normalem]{ulem}
\usepackage{tabularx}
\usepackage{array,multirow}
\usepackage{url}
\usepackage{amssymb}

\DeclareUnicodeCharacter{2212}{-}
\DeclareUnicodeCharacter{03C9}{$\omega$}

\begin{document}

\preprint{AIP/123-QED}

\title{Thermal transport in GaN/AlN HEMTs on 4H-SiC: Role of layer thickness and hetero-interfaces}

\author{Dat Q. Tran}
\email{dat.tran@liu.se}
\affiliation{Center for III-Nitride Technology, C3NiT - Janz\'en and Department of Physics, Chemistry and Biology (IFM), Link\"{o}ping University, SE-58183 Link\"{o}ping, Sweden}
\affiliation{Department of Electrical Engineering, Stanford University, California 94305, USA}

\author{Minho Kim}
\affiliation{Center for III-Nitride Technology, C3NiT - Janz\'en and Department of Physics, Chemistry and Biology (IFM), Link\"{o}ping University, SE-58183 Link\"{o}ping, Sweden}
\affiliation{Wallenberg Initiative Materials Science for Sustainability (WISE), Department of Physics, Chemistry and Biology (IFM), Link\"{o}ping University, SE-58183 Link\"{o}ping, Sweden} 

\author{Okhyun Nam}
\affiliation{Convergence Center for Advanced Nano Semiconductor (CANS), Department of Nano-Semiconductor, Tech University of Korea, 15073 Siheung-si, Korea}

\author{Vanya Darakchieva}
\affiliation{Center for III-Nitride Technology, C3NiT - Janz\'en and Department of Physics, Chemistry and Biology (IFM), Link\"{o}ping University, SE-58183 Link\"{o}ping, Sweden}
\affiliation{Wallenberg Initiative Materials Science for Sustainability (WISE), Department of Physics, Chemistry and Biology (IFM), Link\"{o}ping University, SE-58183 Link\"{o}ping, Sweden}
\affiliation{NanoLund and Solid State Physics, Lund University, 22100 Lund, Sweden}

\author{Plamen P. Paskov}
\affiliation{Center for III-Nitride Technology, C3NiT - Janz\'en and Department of Physics, Chemistry and Biology (IFM), Link\"{o}ping University, SE-58183 Link\"{o}ping, Sweden}

\date{\today}

\begin{abstract}

Thermal transport in high-electron-mobility-transistor (HEMT) structures grown on 4H-SiC substrates by metalorganic-vapour-phase epitaxy (MOCVD) is systematically investigated. The thermal conductivity of the GaN channel and AlN buffer layers is measured by thermoreflectance (TTR). A pronounced thickness dependence of thermal conductivity as a result of phonon-boundary scattering is observed at low temperatures, while this effect becomes significantly weaker at elevated temperatures. The thermal boundary resistance (TBR) at the AlN/4H-SiC and GaN/AlN interfaces is also examined, showing a substantial reduction and eventual saturation with increasing temperature, indicating elastic phonon transport as the dominant mechanism.  
Reliable simulations of the temperature profile across the structures based on the measured thermal metrics highlight the critical role of TBR in thin-channel device and the advantage of thicker channel and buffer layers for efficient heat dissipation in the HEMTs.
\end{abstract}

\maketitle

Lateral AlGaN/GaN high-electron-mobility-transistors (HEMTs) have played a pivotal role in the advancement of GaN technology for high-power (HP) and high-frequency (HF) applications.\cite{Teo_2021, He_2021} The combination of high electron mobility and the large critical electric field (3.3 MV/cm) of GaN enables significantly higher breakdown voltages at a given on-state resistance compared to silicon (Si) and silicon carbide (SiC) devices.\cite{Lidow_2019} As a result, for a specific current rating, smaller die sizes can be used in GaN-based devices, leading to lower power losses and improved efficiency.\cite{Chowdhury_2014,Devi_2022}

HP and HF HEMTs operate at high current and/or high voltage, where Joule heating in the active device region is a severe problem for both performance and reliability. Self-heating effects  have been shown to degrade electron mobility, saturation velocity, breakdown voltage, \cite{Gaska_1998, Turin_2004,Kuzmik_2004} and in extreme cases,  can lead to thermal failure of the device. In  AlGaN/GaN HEMTs, hot spots form locally under the gate region,\cite{Arivazhagan_2021,Ranjan_2019} where the electric field reaches its maximum. Recent studies have reported significant channel temperature rises even at operating powers in the milliwat range,\cite{Raad_2022} underscoring the critical need for effective thermal management in these devices.

A typical AlGaN/GaN HEMT grown on a SiC consists of an AlN buffer layer, a GaN channel layer and an AlGaN barrier.\cite{Pangelly_2012, Chen_2018} For effective thermal management, two parameters are critical: the thermal conductivity (ThC) of the GaN and AlN layers, and the thermal boundary resistance (TBR) at the interfaces. The bulk thermal conductivity of GaN and AlN is relatively high at room temperature, ranging from 200 to 250 W/m.K for GaN\cite{Mion_2006,Shibata_2007,Paskov_2017,Zheng_2019,Tran_2022_PRM} and 270 to 316 W/m.K for AlN.\cite{Rounds_2018,Xu_2019,Cheng_2020,Inyushkin_2020b} However, in thin GaN and AlN layers used in HEMT structures, the ThC is significantly reduced due to enhanced phonon–boundary scattering.\cite{Belkerk_2014, Beechem_2016, Ziade_2017, Tran_2020_APL} The TBR between two dissimilar semiconductor layers primarily arises from mismatches of in their phonon density-of-states (PDOS),\cite{Monachon_2016} but it is also strongly influenced by the quality of the interface and the presence of intrinsic and extrinsic defects. Since thermal transport in HEMT structures occurs over length scales comparable to the phonon mean free path (MFP), the TBR plays a crucial role determining the efficiency of heat dissipation, particularly in the-highfield region near the AlGaN/GaN interface.\cite{Zhang_2023}

In this study, we investigate the thermal transport in AlGaN/GaN HEMT structures grown on 4H-SiC substrates. The thermal conductivity of the GaN and AlN layers is systematically studied as a function of layer thickness and temperature. The TBR at the AlN/4H-SiC and GaN/AlN interfaces is also examined. Finally, we explore how both the layer thermal conductivity and interface TBR influence heat dissipation within the HEMT devices.

Two series of AlGaN/GaN HEMT structures were grown on 4H-SiC substrates using hot-wall metalorganic vapor-phase epitaxy (MOCVD). Both series share the same layer stack of Al$_{0.25}$Ga$_{0.75}$N/GaN/AlN/4H-SiC, from top to bottom as illustrated in Fig. 1(a). The AlGaN barrier has a nominal thickness of 25 nm. In the first series, the AlN buffer layer thickness is fixed at 60 nm, while the GaN channel layer is varied between 150 nm and 2 $\mu$m. In the second series, the GaN channel thickness is kept constant at 150 nm, while the AlN buffer layer thickness varies from 0.12 to 2 $\mu$m.  The edge dislocation densities in the GaN and AlN layers were estimated from the asymmetric ($10\Bar{1}2$) diffraction peak widths and the layer thicknesses were measured using ultraviolet-visible spectroscopic ellipsometry.\cite{Schubert_2025} A summary of the layer thicknesses and the estimated edge dislocation densities (\textit{D}$_E$) for all investigated samples is provided in Table \ref{table1}.

\begin{table}
\caption{\label{table1} Thickness and edge dislocation density of the AlN and GaN layers in the investigated HEMT structures.}
\begin{ruledtabular}
\begin{tabular}{ccccc}
\multirow{2}{*}{Samples} & \multicolumn{2}{c}{Thickness (nm)} & \multicolumn{2}{c}{\textit{D}$_E$ (cm$^{-2}$)} \\
\cline{2-5}
& AlN & GaN & AlN & GaN \\
\hline
S1 & 60 & 2000 & $6.5\times10^9$ & $6.0\times10^8$ \\
S2 & 60 & 150 & $6.5\times10^9$ & $1.0\times10^9$ \\
S3 & 120 & 150 & $6.0\times10^9$ & $1.0\times10^9$ \\
S4 & 500 & 150 & $1.4\times10^9$ & $3.0\times10^9$ \\
S5 & 1000 & 150 & $9.0\times10^8$ & $1.6\times10^9$ \\
S6 & 2000 & 150 & $1.6\times10^9$ & $1.4\times10^8$ \\
\end{tabular}
\end{ruledtabular}
\end{table}

The ThC and TBR was determined by the transient thermoreflectance (TTR) technique, as detailed in our previous reports.\cite{Tran_2022_PRM,Tran_2020_APL, Tran_2020_Physica_B} A gold (Au) transducer layer with a thickness of 200$\pm$5 nm was deposited on top of the samples and the thermoreflectance transients from this layer were measured. The transients were analyzed using a solution of the one-dimensional heat transport equation\cite{Kulish_2001} and the ThC and TBR are extracted from the fitting. Prior to measuring the HEMT structures, the ThC of the 4H-SiC substrate and the TBR at the Au/Al$_{0.25}$Ga$_{0.75}$N interface were independently determined. The rest of the input parameters used in the fitting, i.e. the specific heats of GaN, AlN, 4H-SiC and Au, and the thermal conductivity of Au were taken from literature.\cite{Lee_2011,Touloukian_1970_V1,Touloukian_1970_V4} 

The measured ThCs of the AlN buffer and GaN channel layers as functions of temperature and layer thickness are shown in Figs. 1(b) and 1(c). Across the entire temperature range studied (77-400 K), the ThC of the AlN layers ($k_{\text{AlN}}$) increases with increasing layer thickness (Fig. 1(b)). For the 2 $\mu$m thick layer, $k_{\text{AlN}}$ = 190 W/m.K at $T$ = 300 K is obtained. This value is significantly higher than that reported for AlN layers of the same thickness grown on Si and SiN substrates.\cite{Belkerk_2014} This is likely due to the superior crystal quality (i.e. lower dislocation density) of our AlN layers grown on 4H-SiC. For thinner layers (0.12 $\mu$m and 0.5 $\mu$m), the ThC increases up to $T$ = 200 K, beyond which it saturates. This behavior is attributed to dominant phonon-boundary scattering in thin layers. In contrast, thicker layers exhibit temperature dependence more similar to that of bulk AlN, i. e. the ThC increases to a certain temperature ($T$ = 100 K for 2 $\mu m$ thick layer) and then decreases indicating that three phonon-phonon scattering begins to dominate over phonon-boundary scattering. It should be noted that for bulk AlN the maximum ThC occurs at a lower temperature (about $T$ = 80 K).\cite{Inyushkin_2020b} For comparison Fig. 1(b) also includes the measured ThC of a bulk AlN crystal grown by physical vapor transport (PVT), which follows a temperature dependence of approximately T$^{-1.45}$. At $T$ = 300 K, $k_{\text{AlN}}$ = 340 W/m.K  is obtained in good agreement with previously reported values.\cite{Cheng_2020,Inyushkin_2020b}

Figure 1 (c) presents the ThC of bulk GaN grown by hydride vapor phase epitaxy (HVPE) \cite{Tran_AIP_2023} along with that of 0.15 $\mu$m and 2 $\mu$m  thick layers in the HEMT structures with the 60 nm thick AlN buffer layer. The temperature dependence of the ThC for the GaN layers resembles that observed for the AlN layers. At $T$ = 300 K, the measured ThC, $k_{\text{GaN}}$ is 150 W/m.K for the 2 $\mu$m and 28 W/m.K for the 0.15 $\mu$m thick GaN layer. These values are lower than those of the AlN layers of comparable thickness, which reflects the intrinsically lower ThC of bulk GaN compared to bulk AlN (about $35$\% lower at $T$ = 300 K.) Note that the ThC of bulk GaN is found to follow a temperature dependence of approximately $T^{-1.1}$ . In comparison, the larger exponent observed for AlN is attributed to the stronger contribution of optical phonon scattering, which becomes increasingly significant at elevated temperatures.\cite{Lindsay_2013}

\begin{figure*}
\centering
\includegraphics[scale = 0.54]{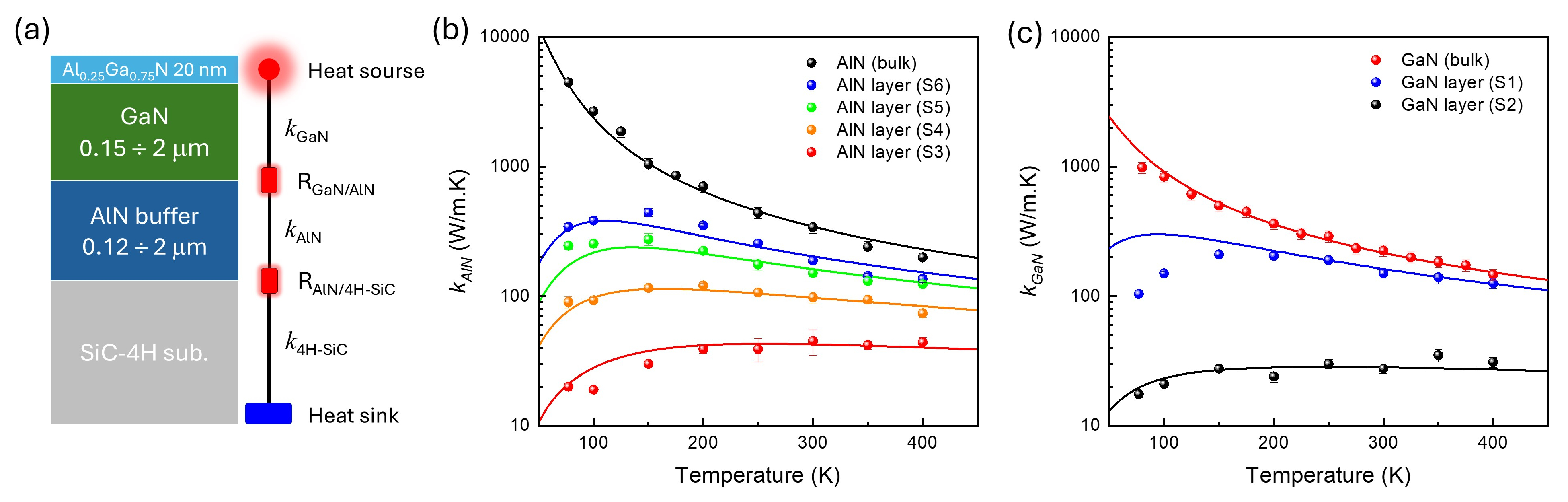}
\caption{(a) Schematic illustration of the HEMT structures grown on a 4H-SiC substrate, along with the thermal propagation path across the layers. (b) and (c) Temperature-dependent thermal conductivity of the AlN buffer and GaN channel layers, respectively. The solid lines represent the calculated thermal conductivity.}
\label{Fig_1}
 \end{figure*}

The solid lines in Figs. 1(b) and 1(c) present the calculated ThC of GaN and AlN. The calculations are performed by a solution of the phonon Bolzman transport equation (BTE) within relaxation time approximation (RTA). In this approach, the thermal conductivity is expressed as \cite{Tran_AIP_2023} 

\begin{equation} \label{eq.1}
    k = \frac{\hbar^2}{6\pi^2k_B T^2} \sum_{\lambda}^{} \int_{0}^{q_{max}} \tau_\lambda(q) n_o(1+n_o) v_{g,\lambda}^2(q)\omega_\lambda^2(q) q^2dq,
\end{equation}
where $k_B$ is the the Boltzmann constant, $\hbar$ is the reduced Planck constant, $T$ is the temperature, $n_o$ is the Bose-Einstein distribution, $\tau$ is the phonon scattering time, and $q$, $\omega$ and $v_g$ are the phonon wavevector, the phonon angular frequency and the phonon group velocity, respectively. The summation is over the all acoustic phonon modes, $\lambda$. The integration is over the first Brillouin zone along the $\Gamma-$A symmetry with ${q_{max}}$ being the phonon wavevector at the zone boundary. The phonon scattering time, $\tau$, is assumed to be additive of all resistive phonon scattering processes via Matthiessen's rule with $ \tau^{-1} = \sum_i \tau_i^{-1}$, where $i$ stands for the type of scattering processes. The three-phonon Umklapp ($U$), the phonon-boundary ($B$), the phonon-isotope ($I$), the phonon-dislocation ($D$) and the phonon-point-defect ($PD$) scattering are considered.

The $U$-scattering is an intrinsic mechanism which governs the thermal conductivity in pristine semiconductor crystals. The scattering rate for this process is given by:\cite{Slack_1964}

\begin{equation}
    \tau_U^{-1} = \frac{\hbar \gamma^2}{Mv_g\theta}(T/\theta)\text{exp}(-\theta/hT) \omega^2,
\end{equation}
where $M$ is the averaged atomic mass, $\theta$ is the Debye temperature and nd $\gamma$ is the Gr\"uneisen parameter for the corresponding acoustic mode. The parameter $h$ governs the temperature dependence of thermal conductivity in defect-free semiconductors and is typically determined by fitting the experimental temperature dependence of thermal conductivity in high-quality bulk crystals. Based on the data analysis in Figs. 1(b) and 1(c), we extracted values of $h = 2.5$ for AlN and $h = 3$ for GaN. The Debye temperatures were calculated from the zone boundary frequencies at A symmetry point of the Brillouin zone $\omega_{max,\lambda}$ by $\theta_{D,\lambda} = \hbar\omega_{max,\lambda}/k_B$.\cite{Morelli_2002} The phonon group velocities and the Gr\"uneisen parameters are  obtained from the calculated phonon dispersion. In our thermal conductivity calculations, mode-averaged  Gr\"uneisen parameters were used: $(\gamma_{TA}, \gamma_{LA}) = (0.18, 1.36)$ for GaN,\cite{Tran_AIP_2023} and $(0.16, 1.04)$ for AlN.\cite{Tran_2023_APL}

In thin layers, $B$-scattering becomes the dominant mechanism governing ThC. The scattering rate is commonly expressed as $\tau_B^{-1} = v_g/L$,\cite{Beechem_2016,Ziade_2017,Morelli_2002} where $v_g$ is the phonon group velocity and $L$ is the layer thickness. However, this simplified treatment tends to underestimate the contribution of high-frequency phonons, leading to a distorted spectral distribution of thermal conductivity ($k$). In reality, only phonons whose bulk mean free paths (MFPs) are comparable to or longer than the layer thickness are significantly impacted by boundary scattering.\cite{Jiang_2016} To more accurately capture this effect, we adopt the following expression for the $B$-scattering rate in this study:

\begin{equation}
 \tau^{-1} = 
\begin{cases}
    v_g/\text{MFP} & \text{if  }  \text{MFP} < L \\
    v_g/L   & \text{if  }  \text{MFP} \geq L
\end{cases}
\end{equation}

The scattering rate expressions for \textit{I}-, \textit{D}- and \textit{PD}-scattering can be found in our previous work.\cite{Tran_2022_PRM} Our analysis has shown that these mechanisms contribute only marginally to the thermal conductivity of the layers investigated here. As shown in Figs. 1(b) and 1(c), there is excellent agreement between the experimentally measured and calculated thermal conductivities of the AlN and GaN layers.

Figure 2(a) shows the temperature dependence of the thermal boundary resistance (TBR), $R_T$, at the AlN/4H-SiC and GaN/AlN interfaces. Experimental values for the TBR ($R_T$) at the AlN/4H-SiC interface are extracted by extrapolating the AlN layer resistance as a function of thickness.\cite{Deng_2019} A larger $R_T$ is obtained at lower temperatures, especially for $T \leq$ 200 K, which is primarily attributed to the suppression of high-frequency phonons at reduced temperatures. Above 200 K, $R_T$ almost saturates, indicating that inelastic phonon scattering processes contribute negligibly in this temperature regime. At room temperature, we obtained $R_T$ = 2.56 $\text{m}^{2}\text{K/GW}$, in good agreement with recently reported value.\cite{Walwil_2025} Due to the limited number of GaN layer thicknesses available in this study, it was not possible to experimentally determine the TBR at the GaN/AlN interface using the above method. Therefore, only the calculated temperature dependence of $R_T$ for the GaN/AlN interface is presented in Fig. 2(a).

\begin{figure*}
\centering
\includegraphics[scale = 0.45]{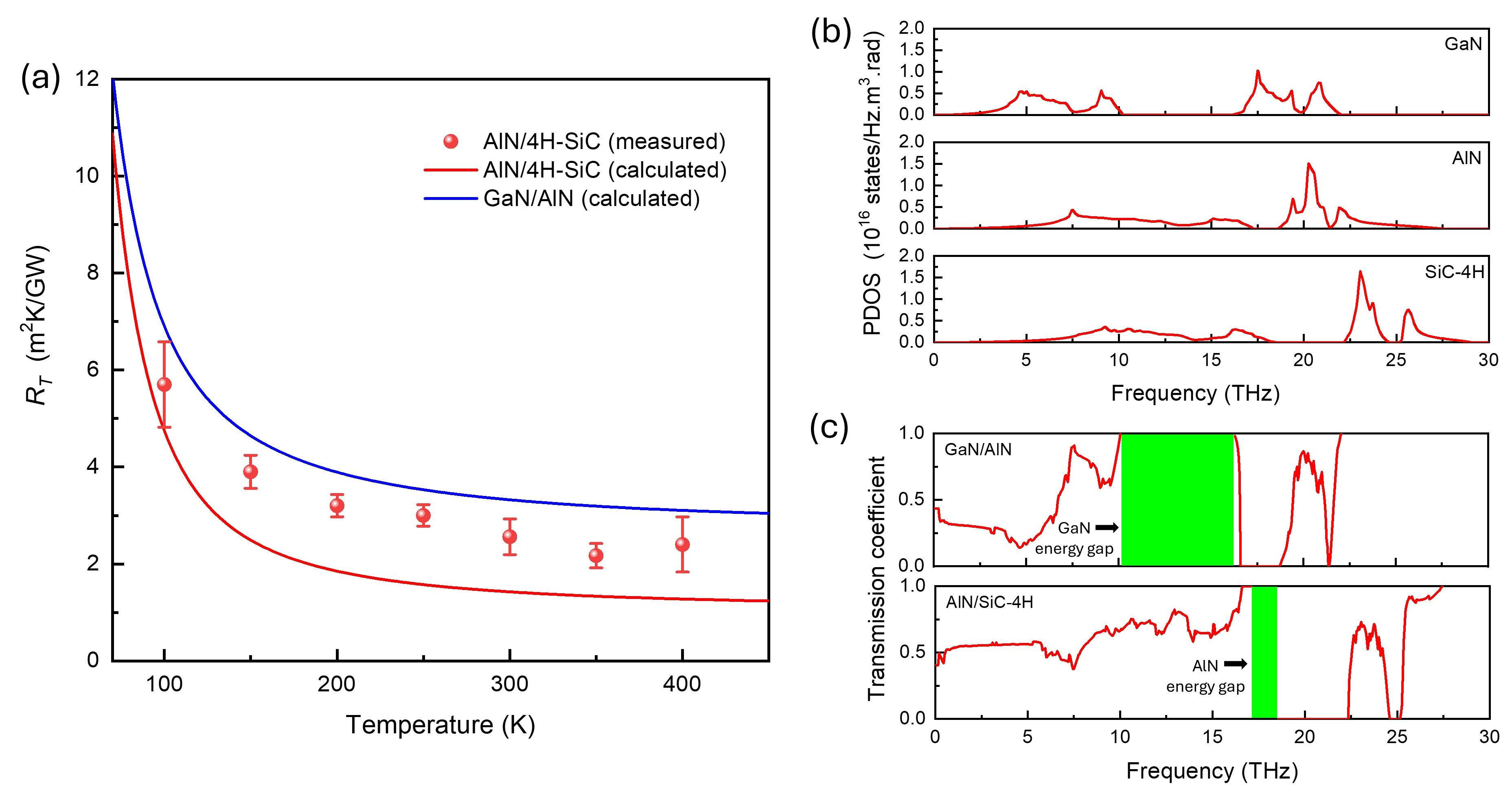}
\caption{(a) Thermal resistance at the AlN/4H-SiC and GaN/AlN interfaces, (b) phonon density of states of GaN, AlN, and 4H-SiC, and (c) phonon transmission coefficients at the AlN/4H-SiC and GaN/AlN interfaces.}
\label{Fig_2}
 \end{figure*}

For the calculation of thermal boundary resistance (TBR), we employ a modified Landauer approach that assumes diffusive phonon transport across the interface between two semiconductors. In this framework, $R_T$ is given by:\cite{Deng_2019}

\begin{equation}
    R_T = \left(\frac{1}{4}\int D(\omega) \frac{\delta n_o(\omega, T)}{\delta T} \overline{v}(\omega) \hbar \omega \overline{\alpha}(\omega) d\omega\right)^{-1}
    \label{eq.4}
\end{equation}
Here, $D$ denotes the PDOS, while $\overline{v}$ and $\overline{\alpha}$ represent the group velocity and the transmission coefficient averaged over the phonon modes. These parameters correspond to the material from which the phonon originates. Under the assumption of diffusive phonon transport (phonon diffusive mismatch model), the average transmission coefficient from material 1 to material 2 is given by: $\overline{\alpha} = D_2 \overline{v}_{2}/(D_1 \overline{v}_{1} + D_2 \overline{v}_{2})$. It should be noted that this expression is valid only under the approximation $\sum_\lambda D_\lambda(\omega)v_\lambda(\omega) = D(\omega)\overline{v}(\omega)$. The values of 
$D$ and $\overline{v}$ were obtained from ab initio calculations using the supercell and finite atomic displacement method, implemented via the Quantum ESPRESSO package\cite{Giannozzi_2009}  and the Phonopy code.\cite{Togo_2023} The calculated PDOS of GaN, AlN, and 4H-SiC along with the corresponding transmission coefficients at the            GaN/AlN and AlN/4H-SiC interfaces are presented in in Fig. \ref{Fig_2}(b) and Fig. \ref{Fig_2}(c), respectively. The results show that the dominant contribution to interfacial thermal conductance arises from phonons with frequencies below the phonon bandgap. Specifically, approximately 75\% of the total thermal conductance is governed by phonons with frequencies in the range of $5 - 10$ THz for the GaN/AlN interface, and $5 - 12$ THz for the AlN/4H-SiC interface.

Across the entire temperature range investigated, the calculated TBR $R_T$ of the AlN/4H-SiC interface is lower than the experimentally measured values, as shown in  Fig. 2(a). This difference could be attributed to the presence of structural defects at the interface, induced by lattice mismatch, which increase the actual TBR in the grown structures.\cite{Chen_2018} The calculated $R_T$ for the GaN/AlN interface is  more than twice as large as that of the AlN/4H-SiC interface. This is primarily due to the greater mismatch in PDOS between GaN and AlN. At room temperature, the value of $R_T$ = 3.3 $\text{m}^{2}\text{K/GW}$ we obtained is in good agreement with previously reported theoretical results.\cite{Polanco_2019,Wang_2021}

To evaluate the temperature distribution across the investigated HEMT structures, we performed numerical calculations of electro-thermal transport using ATLAS TCAD simulator.\cite{ATLAS} A drift-diffusion model was employed for the electrical transport, incorporating both spontaneous and piezoelectric polarization effects to account for the formation of the two-dimensional electron gas (2DEG) at the AlGaN/GaN interface. Self-heating effects were included by introducing a heat source term $H = (\overrightarrow{J_n} + \overrightarrow{J_p})*\overrightarrow{E}$ to the steady-state heat flow equations, where $J$ denotes the current density and $E$ is the electric field generated within the operating devices. A heat sink was introduced by setting a thermal contact with $T = 300$ K at the bottom of the 4H-SiC substrate. The simulations incorporated the temperature-dependent ThC and thermal boundary resistances TBR of all layers, based on both experimental measurements and theoretical calculations described in the previous sections. For the 4H-SiC substrate, out-of-plane ThC of 320 W / mK and in-plane of 470 W / mK were used.\cite{Qian_2017} Simulations were performed for devices with lateral dimensions of 10$\times$10 $\mu$m$^{2}$ with gate-to-source and drain-to-source distances of 3 and 9 $\mu$ m, respectively. The operating bias conditions were set to $V_{GS}$ = 0 V and $V_{DS}$ = 10 V.

\begin{figure}
\centering
\includegraphics[scale = 0.35]{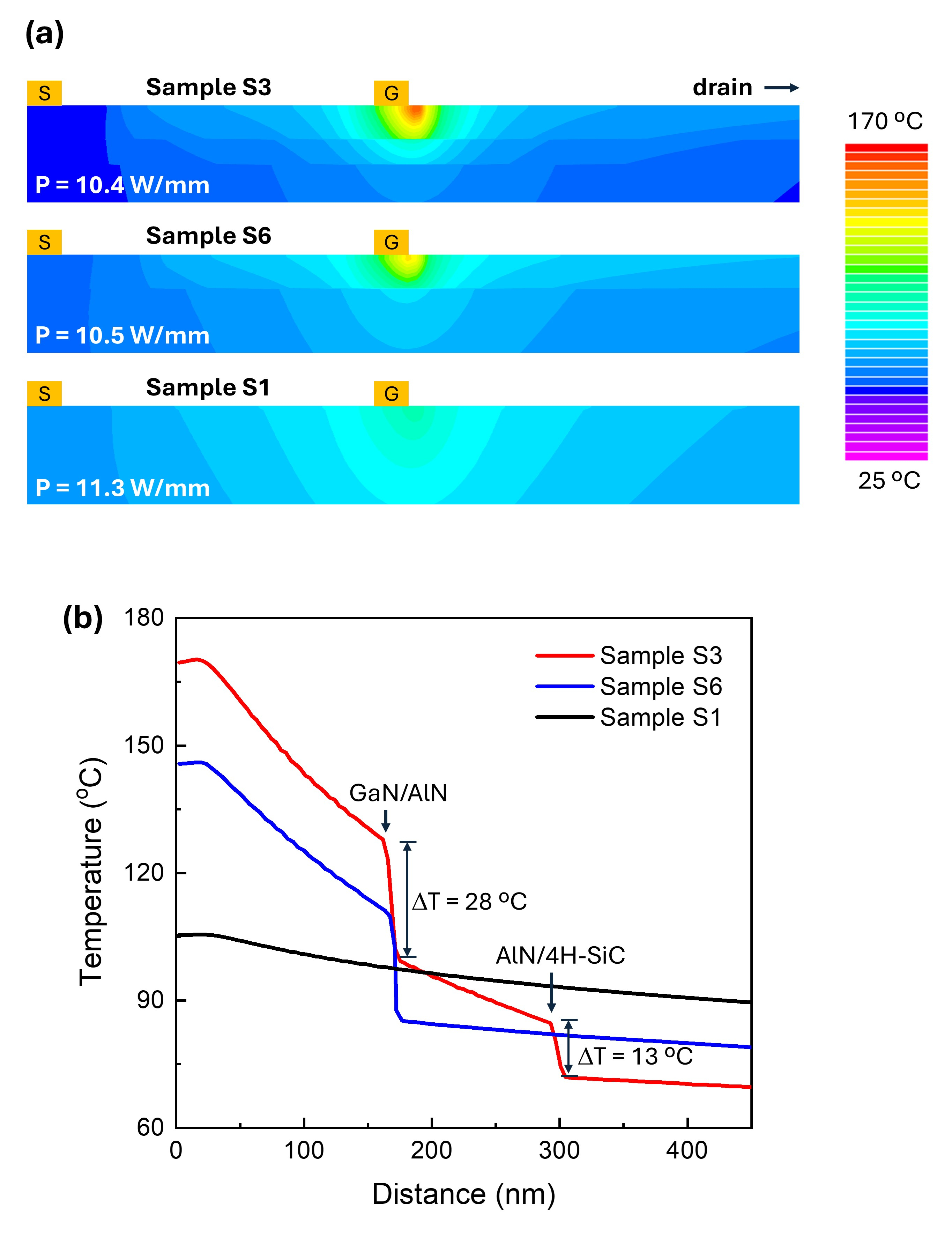}
\caption{(a) Steady-state temperature distribution across the HEMT structures and (b) vertical temperature profiles scanned along the line under hot spot.}
\label{Fig_3}
 \end{figure}

The simulated results for three structures (S1, S3 and S6) are presented in Fig. \ref{Fig_3}. As shown in Fig.\ref{Fig_3}(a), the self-heating induced temperature rise originates beneath the gate, shifts toward the drain side, and spreads across the entire device structure. Compared to the HEMT with thin channel and buffer layers (S3), the hot-spot peak temperature is reduced by 14\% in the structure with the 2 $\mu$m AlN buffer (S6) and by 38\% in the structure with the  2 $\mu$m GaN channel (S1) (Fig. \ref{Fig_3}(b)). This reduction is primarily attributed to the higher thermal conductivity of the thick AlN and GaN layers. The TBR at the GaN/AlN interface plays a significant role in structures S3 and S6, leading to an abrupt temperature drop of 28$^\circ$C. In structure S3, an additional temperature drop of 13$^\circ$C is observed at the AlN/4H-SiC interface. These temperature differences ($\Delta T$) can be described by the relation $Q_{local}\times R_{T}$, where $Q_{local}$ is the local heat flux at the interface and is inversely proportional to the distance from the hot spot at the gate. A larger $\Delta T$ implies a stronger thermal transport barrier. Notably, in structure S1, the interfaces are nearly invisible to the heat source due to their greater distance from the hot spot. This reduced interfacial resistance significantly contributes to the lower peak temperature observed in this structure.

In summary, we present a comprehensive study of the thermal properties of HEMT structures grown on 4H-SiC substrates with varying thicknesses of the GaN channel and AlN buffer layers. The thermal conductivity (ThC) of the AlN buffer and GaN channel layers was measured and calculated as a function of both thickness and temperature. The results show that ThC increases with layer thickness, indicating that phonon-boundary scattering is the dominant limiting mechanism. Thermal boundary resistance (TBR) at the AlN/4H-SiC and GaN/AlN interfaces was also examined. In addition, we performed electro-thermal simulations to model the temperature distribution within the structures. The results reveal that employing a thin GaN channel (150 nm) leads to a significant temperature rise due to both reduced thermal conductivity and the increased impact of TBR at the GaN/AlN interface. In contrast, using a thicker GaN channel (2 $\mu$m) results in a substantial reduction up to 38\% in surface temperature. The use of a thicker AlN buffer layer also contributes to improved heat dissipation, though the effect is less pronounced. This study provides valuable insights into thermal transport mechanisms in GaN-based HEMT structures and offers practical guidance for optimizing device design to mitigate self-heating effects.

\begin{acknowledgments}
This work was performed within the competence center for III Nitride Technology (C3NiT-Janzén) supported by the Swedish Governmental
Agency for Innovation Systems (VINOVA) under the Competence Center Program Grant No. 2022-03139. The authors also acknowledge support from the Swedish Research Council VR (Grant
Nos. 2022-04812 and 2023-04993), Knut and Alice Wallenberg Foundation funded grant 'Transforming ceramics into next generation semiconductors' (Grant No. 2024.0121), the Wallenberg Initiative Materials Science for Sustainability (WISE) funded by the Knut and Alice Wallenberg Foundation and the Swedish Government Strategic Research Area on Functional Materials Materials Science at Linköping University,
Faculty Grant SFO Mat LiU No. 009-00971. First-principles computations were enabled by resources provided by the Swedish National Infrastructure of Computing (SNIC). V.D. acknowledges support by the Knut and Alice Wallenberg Foundation for a Scholar award (Grant No. 2023.0349) and D.Q.T acknowledges support by the Knut and Alice Wallenberg Foundation for a Postdoctoral Fellowship at Stanford University (Grant No. 2023.0492).
\end{acknowledgments}

\section*{DATA AVAILABILITY}
The data that support the findings of this study are available by authors.

\nocite{*}
\bibliography{Manuscript}
\end{document}